\documentclass[a4paper,epsf,12pt]{article}
\usepackage[pctex32]{graphics}
\usepackage{subfigure}
\usepackage{graphicx}

\begin{document}
\topmargin 0pt \oddsidemargin 0mm
\newcommand{\beq}{\begin{equation}}
\newcommand{\eeq}{\end{equation}}
\newcommand{\beqa}{\begin{eqnarray}}
\newcommand{\eeqa}{\end{eqnarray}}
\newcommand{\sr}{\sqrt}
\newcommand{\fr}{\frac}
\newcommand{\mn}{\mu \nu}
\newcommand{\G}{\Gamma}

\begin{titlepage}

\vspace{5mm}
\begin{center}
{\Large \bf Holographic superconductor in the  analytic hairy black
hole } \vspace{12mm}

{\large   Yun Soo Myung \footnote{e-mail
 address: ysmyung@inje.ac.kr} and Chanyong Park \footnote{e-mail address: cyong21@sogang.ac.kr}}
 \\
\vspace{10mm} {\em $^1\,$ Institute of Basic Science and School of
Computer Aided Science \\ Inje University, Gimhae 621-749, Korea \\
$^2\,$ Center for Quantum Spacetime (CQUeST), Sogang University, Seoul 121-742, Korea }

\end{center}
\vspace{5mm} \centerline{{\bf{Abstract}}}
 \vspace{5mm}
We study the charged black hole of hyperbolic horizon  with scalar
hair (charged Martinez-Troncoso-Zanelli: CMTZ black hole) as a model
of analytic hairy black hole for holographic superconductor.  For
this purpose, we investigate the second order phase transition
between CMTZ and hyperbolic Reissner-Nordstr\"om-AdS (HRNAdS) black
holes. However, this transition unlikely occur.  As an analytic
treatment for holographic superconductor, we develop superconductor
in the bulk and superfluidity on the boundary using the CMTZ black
hole below the critical temperature. The presence of charge destroys
the condensates around the zero temperature, which is in accord with
the thermodynamic analysis of the CMTZ black hole.
\end{titlepage}
\newpage
\renewcommand{\thefootnote}{\arabic{footnote}}
\setcounter{footnote}{0} \setcounter{page}{2}

\section{Introduction}
Recently, the AdS/CFT correspondence has widely applied to
investigating the superconductivity. In order to describe a
superconductor, we introduce the temperature  and  charge density in
the boundary field theory by
 adding a charged black hole~\cite{Gubs,HHH}. Then, we  make a condensate through coupling of a charged
 scalar field $\psi$ to a Maxwell field to define  Cooper pair's operator and its charge density in the boundary theory.
   In these cases, the charged scalar field is minimally coupled to gravity.
 To represent the phase diagram for
 holographic superconductor, one supposes  a system which admits black hole with
 scalar hair at low temperature, while  at
 high temperature  there is a black hole without hair, namely Reissner-Nordst\"om-AdS$_4$ (FRNAdS) black hole with flat horizon.
   In all of these gravity constructions~\cite{HartC}, the onset of the phase transition
 is triggered by a zero mode for the field that gives hair to the black hole.
 The FRNAdS black hole is  unstable against the perturbation of
 the charged scalar  because the effective mass $m^2_{\rm \psi}=m^2+p^2g^{tt}A_t^2$ of the scalar field
 becomes too negative~\cite{Horo}.  Then, this perturbed scalar can  develop a nontrivial
 profile without a $\psi^4$-term to stabilize the run away direction
 of $\psi$.
 Superfluid phase transitions are associated with spontaneous symmetry breaking,
 while  superconducting phase transitions are triggered by the Abelian-Higgs mechanism.
 Hence, superfluidity appears in the boundary field theory and
 superconductivity in the bulk gravity~\cite{HerzJ}. In terms of AdS/CFT picture,
 it implies that an instability of FRNAdS black hole to develop
 scalar hair is dual to a superfluid  phase
 transition. Noting that the AdS/CFT correspondence maps classical
 gravity to a strongly interacting field theory, the correspondence
 opens a window onto  strongly interacting superconductors and superfluids where
the known BCS theory and weak coupling techniques are
inapplicable~\cite{Horo}. Also, it is suggested  that the endpoint
of instability of RNAdS black hole with spherical horizon  is a
numerical hairy black hole~\cite{MFK}.

In this direction, most of studies were being performed numerically.
Solving the equations on the gravity side reduces to a set of
nonlinear, coupled differential equations which could be solved on a
computer. We call these as {\it numerical hairy black hole
solutions}. This is so called  because we could not obtain {\it an
analytic hairy black hole solution} except a few cases: a
complicated $p$-wave superconductor~\cite{BHMS,HPu} and  a gapless
superconductor of neutral MTZ black hole~\cite{KPS2}. In the former,
the dual boundary field theory has a vector rather than a scalar
order parameter. On the other hand, in the latter, the charged
scalar $\phi$  is conformally coupled to gravity and the scalar
potential plays an essential  role to develop scalar condensation. A
second order phase transition between topological black hole and MTZ
black hole~\cite{MTZ} occurred at the critical
temperature~\cite{Siop,Myungsh}. It implies that below the critical
temperature, the MTZ black hole acquires its scalar hair, while an
electromagnetic perturbation determines the conductivity and then,
the superfluid density of the boundary theory.

At this stage, we would like to mention the difference on the
mechanism of condensation of the scalar field. In the case of a
FRNAdS black hole, the scalar field condensates because the
Abelian-Higgs mechanism was operating  without
$\psi^4$-term~\cite{HerzA}, while for the MTZ black hole, the
condensation of the scalar was made by conformally coupling to
gravity and  the presence of $\phi^4$-term in the potential. The
conformal coupling provides a single effective mass $m_{\rm
\phi}^2=-1/l^2$, which is  larger than the Breitenlohner-Freedman
bound, $m^2_{\rm BF}=-9/4l^2$ in the AdS$_4$ spacetime~\cite{BFB}.
Importantly, the presence of $\phi^4$-term is essential for
developing  an exact scalar hair in the hyperbolic
horizon~\cite{CMTZ}.  It is worth to note that the presence of
potential and horizon topology gives rise to an exact scalar hair on
the gravity side.  Hence, it is considered as an exact gravity dual
of a gapless superconductor~\cite{KPS2}. However, the MTZ black hole
is considered as a probe limit  because it belongs to a neutral
black hole.

Fortunately, the charged MTZ (CMTZ) black hole  was found with
scalar hair when the coefficient $\alpha$ of $\phi^4$-term is
between 0 and $2\pi G/3l^2$~\cite{CMTZ}, where the upper bound
corresponds to the MTZ black hole~\cite{MTZ} and the lower bound is
not allowed. Hence, it is very interesting to investigate whether or
not a pair of the CMTZ and HRNAdS black holes provides a second
order phase transition.  It was proposed that this transition is
likely possible to occur for a special case of $\tilde{q}^2=-G\mu^2$
HRNAdS black hole~\cite{KPS1}.  If this pair is  the case, it would
provide a really analytic  gravity dual to a holographic
superconductor. However, the previous analysis might be wrong
because the $\tilde{q}^2=-G\mu^2$ case
 is not allowed in the HRNAdS black holes.  Moreover, the entropy of
the CMTZ black hole is not the Bekenstein-Hawking entropy but it
should be obtained using either the Euclidean action approach or the
Wald method because of the conformal coupling of scalar to gravity.
In this case,  the negative entropy appears~\cite{CMTZ}, which in
turn confines the free energy to restricted ranges, depending on the
charge  $q$. Therefore, the presumed second order phase transition
unlikely occurs between CMTZ and HRNAdS black holes.

On the other hand, we have an analytic charged black hole with
scalar hair, even though the corresponding  black hole without
scalar hair is not clearly known. In this case, we would like to
make a progress on the superfluidity on the boundary by studying the
superconductivity on the gravity side.   This situation is opposite
to a case of numerical hairy black holes  that the charged black
holes without scalar hair  were known explicitly, while the charged
black holes with scalar hair were found numerically.

 The
organization of this work is as follows. Section 2 is devoted to
reviewing  the hyperbolic Reissner-Nordstr\"om-AdS$_4$ (HRNAdS)
black hole as charged black hole without scalar hair.  We study the
CMTZ black hole as the charged black hole with scalar hair in
section 3.   We investigate superconductivity and superfluidity for
the CMTZ black hole thoroughly  in section 4. Finally, we discuss
our results by comparing the known results.

\section{HRNAdS black holes}
Topological black holes in asymptotically anti-de Sitter spacetimes
were first found in three and four dimensions~\cite{Lemos}. Their
black hole horizons are Einstein spaces of spherical, hyperbolic,
and flat  curvature for higher dimensions more than
three~\cite{Vanzo,BLP}. The standard equilibrium and off-equilibrium
thermodynamic approaches to these black holes are possible to show
that they are treated as the extended thermodynamic systems, even
though their horizons are not spherical.

We start with the Einstein-Maxwell action \beq I_4[g,A_\mu]= \int
d^4x\sqrt{-g}\Bigg[\frac{R-2\Lambda}{16 \pi
G}-\frac{1}{16\pi}F^{\mu\nu}F_{\mu\nu}\Bigg] \label{4SDS} \eeq where
$\Lambda=-3/l^2$ with $l$ is the curvature radius of AdS$_4$
spacetimes. For $A=0$ case, we obtain  the action for the four
dimensional topological black hole. For
$A=-\frac{\tilde{q}}{\rho}dt$, the charged topological black hole
(CTBH) in  AdS$_4$ spacetimes are given by
 \beq ds^{2}_{\rm CTBH}=g_{\mu\nu}dx^{\mu}dx^{\nu}=-f_k(\rho)dt^2 +\fr{1}{f_k(\rho)}d\rho^2+\rho^2d\Sigma_k^2, \label{4DBMT} \eeq
 where the metric function $f_k(\rho)$ is given by
\beq f_k(\rho)=k+\fr{ \rho^2}{l^2}-\fr{2G\mu}{\rho}+ \fr{
G\tilde{q}^2}{\rho^2} \eeq where $\mu$ and $\tilde{q}$ are related
to the mass parameter and charge of black hole, respectively. For
$\tilde{q}=0$, the above reduces to topological black hole (TBH).
For $k=-1$ and $\tilde{q}^2=-G\mu^2$, the metric function leads to
the same form  as (\ref{cmtzmetric}) for the CMTZ black hole.
However, this case is not allowed because $\tilde{q}^2=-G\mu^2$
implies an imaginary charge ($\tilde{q}=i \sqrt{G}\mu$).
$d\Sigma_k^2$ describes the 2D horizon geometry with a constant
curvature as Einstein space \beq d\Sigma_k^2= d\theta^2+
p^2_k(\theta) d\varphi^2 \equiv h^k_{ij}dx^idx^j, \eeq where
$p_k(\theta)$ is given by \beq p_{0}(\theta) =\theta, ~p_{1}(\theta)
=\sin \theta, ~p_{-1}(\theta) =\sinh \theta. \eeq
 For $k=-1$, one has
that $\theta \ge 0$ and $0 \le \varphi \le 2\pi$ denote the
coordinates of the hyperbolic space $H^2$. Here we define $k$=1,~0,
and $-1$ cases as the Reissner-Norstr\"om-AdS black hole (RNAdS),
flat RNAdS black hole (FRNAdS), and  hyperbolic RNAdS (HRNAdS) black
holes~\cite{myungsds}, respectively.  It is easy to check that the
metric (\ref{4DBMT}) satisfies \beq R=4\Lambda, \eeq when the
horizon is an Einstein space \beq R_{ij}=k h^k_{ij}. \eeq In this
work, we are interested in the negative curvature with $k=-1$ only.
Then, the horizon space is a hyperbolic manifold of
$\Sigma_{k=-1}=H^2/\Gamma$, where $H^2$ is 2D hyperbolic space and
$\Gamma$ is a freely acting discrete subgroup of the isometry group
$O(2,1)$ of $H^2$~\cite{BM}. This manifold has genus $g\ge 2$ and
its area is given by $\sigma=4\pi(g-1)$. For numerical calculations,
we choose $g=2$, $\sigma=4\pi$, $G=1$, and $l=1$. The boundary has
the metric \begin{equation} ds^2_\partial =-dt^2+l^2 d\sigma^2
\end{equation}
which is a hyperbolic manifold of radius $l$ and its curvature is
negative as $-1/l$.

\section{CMTZ black holes}

In order to understand the role of scalar and its potential well, it
would be better to start with the conformally coupled
action~\cite{KPS2} \beq I_4[g,\phi,A_\mu]= \int
d^4x\sqrt{-g}\Bigg[\frac{R-2\Lambda}{16 \pi
G}-(D_\mu\phi)^*(D^\mu\phi)-\frac{1}{6}R \phi\phi^*-4\alpha (\phi
\phi^*)^2-\frac{1}{16\pi}F^{\mu\nu}F_{\mu\nu}\Bigg] \label{4SDSc},
\eeq where a conformally charged coupled scalar $\phi$ appears with
a conventional potential term $\alpha (\phi\phi^*)^2$  and
\begin{equation}
D_\mu\phi\equiv \partial_\mu\phi+ipA_\mu \phi.\end{equation} We note
the mass dimensions of $[\phi]=[\phi^*]=[A_\mu]=1$. For $\phi=0$,
one has the action (\ref{4SDS}) for HRNAdS black hole.  We note that
$\alpha$ is an arbitrary coupling constant, but it was given by
$\alpha=2\pi G/3l^2$ for the MTZ black hole without Maxwell
field~\cite{MTZ}. The equations of motion are given by
 \begin{eqnarray}
\label{eq11}&& G_{\mu\nu}+\Lambda  g_{\mu\nu}=8 \pi G
\Big(T^\phi_{\mu\nu}+
 T^{em}_{\mu\nu}\Big), \\
\label{eq12}&& \frac{1}{\sqrt{-g}}D_\mu(\sqrt{-g} D^\mu\phi)=\frac{1}{6}R\phi +8\alpha \phi^2\phi^*, \\
\label{eq13}&& \nabla_\nu F^{\mu\nu}=4\pi J^\mu,
 \end{eqnarray}
with
\begin{eqnarray}
 T^{\phi}_{\mu\nu}&=&\Big[D_\mu\phi(D_\nu\phi)^*+D_\nu\phi(D_\mu\phi)^*\Big]-g_{\mu\nu}(D_\beta\phi)^*(D^\beta\phi)
 \\ \nonumber
  &+&\frac{1}{3}\Big[g_{\mu\nu}\nabla^2-\nabla_\mu\nabla_\nu +G_{\mu\nu}\Big]\phi\phi^* -4\alpha (\phi\phi^*)^2g_{\mu\nu},
 \end{eqnarray}
 and
 \begin{equation}
 T^{em}_{\mu\nu}= \frac{1}{4\pi}
\Big(F_{\mu\beta}F_{\nu}~^{\beta}-\frac{1}{4}
g_{\mu\nu}F^2\Big),~J_\mu=-ip \Big(\phi^*D_\mu \phi-{\rm c.c}\Big).
\end{equation} Since  the scalar field is conformally coupled,
the total energy-momentum tensor is traceless.  The Ricci
 curvature scalar is constant
\begin{equation}
\bar{R}=4\Lambda.
\end{equation}
 Since Eq. (\ref{eq12}) can be rewritten as
\begin{equation}
\frac{1}{\sqrt{-g}}D_\mu(\sqrt{-g} D^\mu\phi)=\frac{\partial V_{\rm
eff}}{\partial \phi^*},
\end{equation}
it implies the form of an effective potential
\begin{equation}
V_{\rm eff}(\phi)=\frac{1}{6}\bar{R}\phi\phi^*+4\alpha
(\phi\phi^*)^2.
\end{equation}
\begin{figure}[t!]
   \centering
   \includegraphics{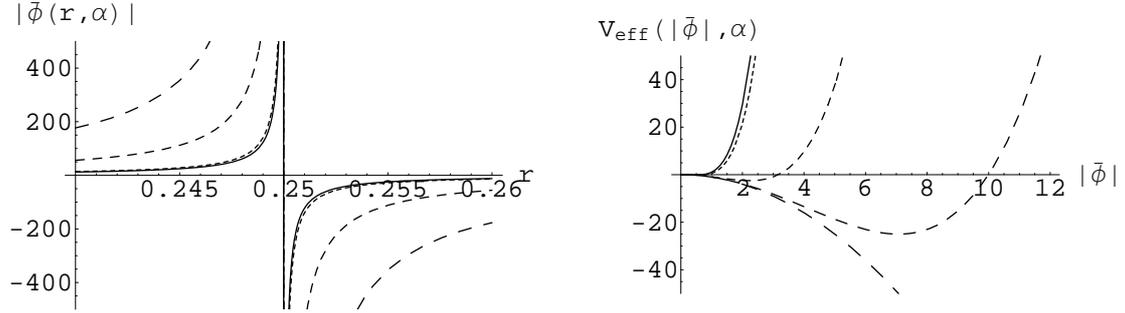}
\caption{ Modulus of scalar $|\bar{\phi}(r,\alpha,\mu)|$ with
$\mu=-\frac{1}{4}$ for  $\alpha=2\pi G/3,1.6,0.1,0.01$ and its
effective potentials $V_{eff}(|\bar{\phi}|,\alpha)$ for $\alpha=2\pi
G/3,1.6,0.1,0.01,0$ from top to bottom. We observe that a scalar
singularity at $r=-G\mu=1/4$ exists for all $\alpha$, which is
responsible for scalar hair in the CMTZ black holes. } \label{fig.1}
\end{figure}
Taking the scalar to be  the modulus
\begin{equation}
\phi \to \frac{|\phi|}{\sqrt{2}} \end{equation}
 together with $p \simeq 0$\footnote{Precisely, for $p\not=0$, the CMTZ black hole solution is not an exact solution to
 (\ref{eq11})-(\ref{eq13}). Actually, we could not obtain the exact complex scalar hairy black hole including the nonzero $p$.
In this work, we use the CMTZ black hole to investigate the density
dependence of the dual condensed matter system in the small $p$
limit. }, all equations (\ref{eq11})-(\ref{eq13}) reduce to those in
the neutral scalar coupled system. Then, the solution is
  given by the CMTZ black hole \beq \label{metricf}
ds^2_{M}=-f_{M}(r)dt^2+ \frac{dr^2}{f_M(r)}+r^2d\Sigma_{k=-1}^2\eeq
where the metric function $f_M(r)$ is given by  \beq
\label{cmtzmetric}
f_M(r)=\frac{r^2}{l^2}-\Bigg(1+\frac{G\mu}{r}\Bigg)^2. \eeq The
modulus of scalar field and the electromagnetic potential take the
forms, respectively, \beq \label{solphi} |\bar{\phi}(r,\alpha,\mu)|=
- \sqrt{\frac{1}{2 \alpha
l^2}}\frac{G\mu}{r+G\mu},~~\bar{A}_t=-\frac{q}{r}. \eeq As is shown
in Fig. 1, the modulus of scalar  and its potential are depicted for
$\alpha=\frac{2\pi G}{3l^2}, 1.6, 0.1, 0.01,0$. The modulus of
scalar field  has a simple pole at $r=-G\mu$ for all $\alpha$ only
if $\mu$ is negative, while shapes of its potential are drastically
changed from $\alpha=0$, $V_{\rm eff}(\phi)=-|\bar{\phi}|^2/l^2$ to
$\alpha \not=0$ case of $V_{\rm eff}=-|\bar{\phi}|^2/l^2+\alpha
|\bar{\phi}|^4$. This makes a difference between $\alpha=0$ and
$\alpha\not=0$. $\alpha=\frac{2\pi G}{3l^2}$ corresponds to the
neutral MTZ black hole. We note here that the singularity of modulus
at $r=-G\mu$ is essential for having the CMTZ black hole dressed by
scalar field and electric charge.

 Importantly, the charge is no longer an
independent parameter and thus it should be determined by the
relation
\begin{equation} \label{qmur}
q^2=G\mu^2\Bigg(-1+ \frac{2\pi G}{3\alpha l^2}\Bigg)
\end{equation}
which determines, from the condition of $q^2\ge 0$, the lower and
upper bound for the coupling constant $\alpha$
  as
\begin{equation} \label{allowedR}
0<\alpha \le  \frac{2\pi G}{3l^2}.
\end{equation}
Because of the above bound, it is convenient to  express $\alpha$ in
terms of $\mu$ and $q$
\begin{equation} \label{newalpha}
\alpha(\mu,q)=\frac{2 \pi G^2 \mu^2}{ 3l^2(G\mu^2+q^2)}
\end{equation} which means that ``$\mu$ and $q$" are two conserved
quantities for the CMTZ black hole.

In the case of positive mass $\mu
>0$, analytic hairy black holes possess a single event horizon located
at
\begin{equation}
r_+=\frac{l}{2} \Bigg(1+ \sqrt{1+ \frac{4G \mu}{l}}\Bigg)
\end{equation}
which satisfies $r_+>l$ and the scalar is regular everywhere.
However, this case  has nothing to do with developing holographic
superconductor. On the other hand, an important feature of the hairy
black hole is that  a complicated horizon structure appears for
$-l/4G<\mu<0$. The hairy black hole has three horizons located at
\begin{equation}
r_\pm=\frac{l}{2} \Bigg(1\pm \sqrt{1+ \frac{4G
\mu}{l}}\Bigg),~r_{--}=\frac{l}{2} \Bigg(-1+ \sqrt{1- \frac{4G
\mu}{l}}\Bigg)
\end{equation}
which satisfy an inequality
\begin{equation} \label{Ineq}
0<r_{--}<-G\mu<r_-<\frac{l}{2}<r_+.
\end{equation}
The outermost $r_+$ and innermost $r_{--}$ horizons are event
horizons and thus one may say that this metric describes ``a black
hole in the  black hole". The causal structure of Penrose diagram
was constructed in Ref.\cite{CMTZ}.

 Thermodynamic quantities of the CMTZ are given by
 Hawking temperature $T_M=\frac{f'_M(r_+)}{4\pi}$, mass $M_M=\frac{\sigma}{4
\pi}\mu$, heat capacity $C_M=\Big(\frac{\partial M_M}{\partial
T_M}\Big)_Q$ by~\cite{MTZ,Siop,Myungsh} \begin{equation}
\label{MBH}T_M(r_+)=\frac{1}{2\pi
l}\Big[\frac{2r_+}{l}-1\Big],~M_M(r_+)=\frac{\sigma r_+}{4\pi
G}\Big(\frac{r_+}{l}-1\Big),~ C_M(r_+)= \frac{\sigma l^2}{4
G}\Big(\frac{2r_+}{l}-1\Big), \end{equation} which  are  the same
quantities as in the MTZ black hole. Here we observe a linear
relation  between temperature and heat capacity
\begin{equation} \label{TCM}
T_M=\frac{2G}{\pi \sigma l^3} C_M =\frac{2G}{\pi \sigma l^3}S_{\rm
MTZ}
\end{equation}
with the entropy for the MTZ black hole \begin{equation}
\label{mtzent} S_{\rm MTZ}(r_+)=\frac{\sigma l^2}{4
G}\Big(\frac{2r_+}{l}-1\Big).
\end{equation}
However, since the conformal coupling term appears in the action,
the entropy should be obtained by making use of the Euclidean action
approach~\cite{CMTZ} and the Wald method~\cite{NVZ} as
\begin{equation} \label{entropy}
S_M(r_+,q)=\frac{\sigma r_+^2}{4 \tilde{G}}=\frac{\sigma l^2}{4 G}
\Bigg(\frac{2r_+}{l}-1-\frac{Gq^2}{r_+^2}\Bigg),
\end{equation}
where an effective Newtonian constant at the event horizon is given
by \begin{equation} \frac{1}{\tilde{G}}=\frac{1}{G}
\Big(1-\frac{4\pi G}{3}|\bar{\phi}|^2\Big). \end{equation} The
positivity of entropy $S_M \ge 0$ requires a bound
\begin{equation} \label{bound}
r_+^2\Big(\frac{2r_+}{l}-1\Big) \ge  Gq^2.
\end{equation}
It requires lower bounds for being the positive entropy.  As is
depicted in Fig. 2, the allowed region depends on the charge  $q$
critically.  The MTZ black hole corresponds to  the largest region
between $l/2 \le r_+ <\infty$. However, the permitted region gets
narrow and narrow as the charge $q $ increases. As $q \to \infty$,
the allowed region approaches  a point at $r_+=\infty$ and thus, we
could not define the entropy.
 We note that the first law
of thermodynamics $dM_M=T_MdS_M-\Phi dQ$  with $\Phi=-q/r_+$ and
$Q=\sigma q/4\pi$ is no longer satisfied for the CMTZ black hole, in
contrast to the MTZ black hole with $q=0$.
\begin{figure}[t!]
   \centering
\includegraphics{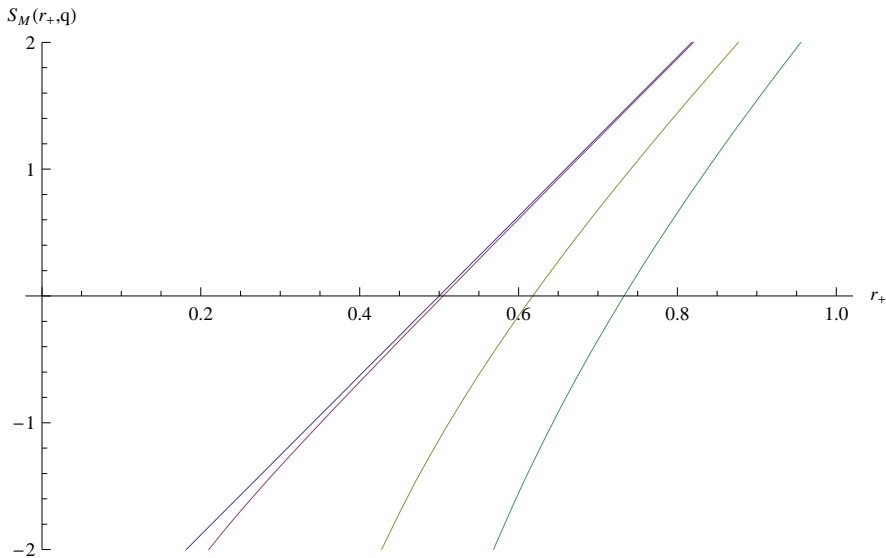}
\caption{Entropy graph: the allowed regions for positive entropy
$S_M(r_+,q)$ with $q=0,~0.05,~0.3,~0.5$ from top to bottom.
 The first case (top) corresponds to the  MTZ black hole with $q=0$: $0.5 \le r_+ <\infty$.
The permitted  regions  gets narrow and narrow as $q$ increases from
0.05 to 0.5: $0.51\le r_+<\infty$ for $q=0.05$, $0.62 \le r_+<\infty
$ for $q=0.3$, and $0.73 \le r_+<\infty $ for $q=0.5$ (bottom). }
\label{fig.2}
\end{figure}

We would like to mention the thermodynamic property of extremal CMTZ
black hole since it may describe the zero temperature limit of the
holographic superconductivity~\cite{Horo}.
 We note that the temperature and heat capacity at the
extremal point $r_+=r_e=l/2$ take the form
\begin{equation}
T_M^e=C_M^e=0.
\end{equation}
The entropy for extremal CMTZ black hole is always  negative as is
shown by
\begin{equation} \label{extremalE}
S^e_M\Big(\frac{l}{2},q\Big)=-\sigma q^2 \le 0
\end{equation}
for  $q\ge 0$ in (\ref{allowedR}). The lower equality represents the
zero entropy of extremal MTZ black hole. Hence, all allowed region
for $r_+$ depending $q$ exclude their extremal black holes at
$r=r_e$.

 The free energy is a key
quantity to see the second order phase transition between HRNAdS and
CMTZ black holes. Actually, to study the first order phase
transition as the Hawking-Page transition~\cite{HP}, one requires
the heat capacity and free energy because it describes a phase
transition from  a hot gas, via small unstable black hole with
negative heat capacity, to large stable black hole with positive
heat capacity in AdS$_4$ spacetimes~\cite{MKP}. However, the second
order phase transition occurs  between two different black holes
with positive heat capacity. Thus, it suggests that the heat
capacity does not play the role in determining the second order
phase transition.

In order to see whether the second order phase transition occurs
 between HRNAdS and CMTZ
black holes, we note that the free energy  for CMTZ black hole
defined by
\begin{equation} F_M= M_M-T_MS_M+\Phi Q
\end{equation}
 takes a   form
 \begin{equation} \label{freeenergy}
 F_M(r_+,q)=\frac{\sigma r_+}{4\pi
G}\Big(\frac{r_+}{l}-1\Big)-\frac{\sigma(2r_+-l)}{8\pi
G}\Bigg(\frac{2r_+}{l}-1\Bigg) -\frac{\sigma l q^2}{8 \pi r_+^2}
\nonumber.
\end{equation}
Here we easily check that for $q=0$, $F_M$ reproduces the free
energy for the MTZ black hole
 \begin{equation} \label{MTZF} F_{\rm MTZ}(r_+)=-\frac{\sigma}{8\pi
 G}\Big(\frac{2r_+^2}{l}-2r_+ +l\Big). \end{equation}
In order to implement the phase transition to the corresponding
black hole without scalar hair, we need the free energy without any
restriction. However, it is clear that the bound (\ref{bound}) puts
restriction on the free energy.  The onset of black hole charge
disturbs a transition to the corresponding black hole without scalar
hair.  A promising  case is the probe limit that  a pair of neutral
black holes of MTZ and TBH provides the second order phase
transition~\cite{Siop,Myungsh}.

Finally, we argue that  a  $q$-dependent free energy  is not
suitable (sufficient) for studying a transition to the corresponding
black hole without scalar hair. In the next section, we will show
explicitly that the onset of charge density on the boundary disturbs
making a solid superconducting and superfluidity phase.

\section{ Gapless superconductivity}

We remind that an analytic hairy black hole was known. Even though
the corresponding black hole without scalar hair is unknown, we wish
to explore  the superfluidity on the boundary by studying the
superconductivity on the gravity side.   This situation is opposite
to the numerical hairy black hole case that  the charged black hole
without scalar hair is known, whereas  the charged black hole with
scalar hair is found numerically. In that case, an instability of
scalar field near (at) the event horizon develops the scalar hair
and thus, the superfluidity on the boundary.

Below the critical temperature defined by $T_0 =  \frac{1}{2 \pi
l}$, the CMTZ black hole solution acquires a scalar hair and thus, a
condensation forms. Near the boundary, the modulus of scalar field
could be expanded as \beq |\bar{\phi}| = \frac{\phi^{(1)}}{r} +
\frac{\phi^{(2)}}{r^2} + \cdots , \eeq where \beq \label{coeffcond}
\phi^{(1)} = - \sqrt{\frac{1}{2 \alpha l^2}} \ G \mu \qquad {\rm and
} \qquad \phi^{(2)} = \sqrt{\frac{1}{2 \alpha l^2}} \ G^2 \mu^2.
\eeq Two non-vanishing coefficients correspond to the condensates of
two dual scalar operator ${\cal O}_i$ $(i=1,2)$, respectively, \beq
\left< {\cal O}_i \right> = \sqrt{2} \phi^{(i)} \ , \ \ i = 1,2 \ .
\eeq We emphasize that unlike  the numerical hairy black hole, the
existence of two condensates does not always imply  an instability
of CMTZ black hole. As was in (\ref{MBH}), one notes that  this
black hole is thermodynamically stable when $\alpha$ approaches its
neutral value ($\alpha \to \frac{2\pi G}{3l^2}$, equivalently $q\to
0$) because $C_M
>0$ and $F_M(r_+,q \to 0)=F_{\rm MTZ}<0$ in (\ref{MTZF}).
We express $\alpha$ in (\ref{newalpha}) in terms of $q$ and  \beq
\mu(T_M) = - \frac{T_0^2 - T_M^2}{8 \pi G T_0^3} \eeq
 through
$r_+=\pi l^2(T_M+T_0)$.
 Furthermore,  substituting  these into
(\ref{coeffcond}) leads  to  two condensates which  take the forms
\begin{eqnarray}    \label{twocond1}
\left< {\cal O}_1 \right> &=& \sqrt{\frac{3}{2 \pi}} \ \sqrt{ q^2 +
\frac{(T_0^2 - T_M^2)^2}{64 \pi^2 G  T_0^6}},   \\
\label{twocond2}\left< {\cal O}_2 \right> &=& \frac{\sqrt{3} (T_0^2
- T_M^2 )}{8 \sqrt{2} \ \pi^{3/2} T_0^3} \ \sqrt{ q^2 + \frac{(T_0^2
- T_M^2)^2}{64 \pi^2 G  T_0^6}} ,
\end{eqnarray}
which for $q=0$, reduce to the same expressions found  in Ref.
\cite{KPS2}.   When $q$ is turned on, the first condensate $\left<
{\cal O}_1 \right>$ does not go to zero at the critical temperature
$T_M = T_0$.  Since the first condensate  does not become zero at
the critical temperature, unlike the MTZ black hole, the second
order phase transition between HRNAdS and  CMTZ black holes may not
occur. This is consistent with the thermodynamic  analysis of the
black hole in the previous section.
\begin{figure}[t!]
  \centering
   \includegraphics[angle=0,width=0.7\textwidth]{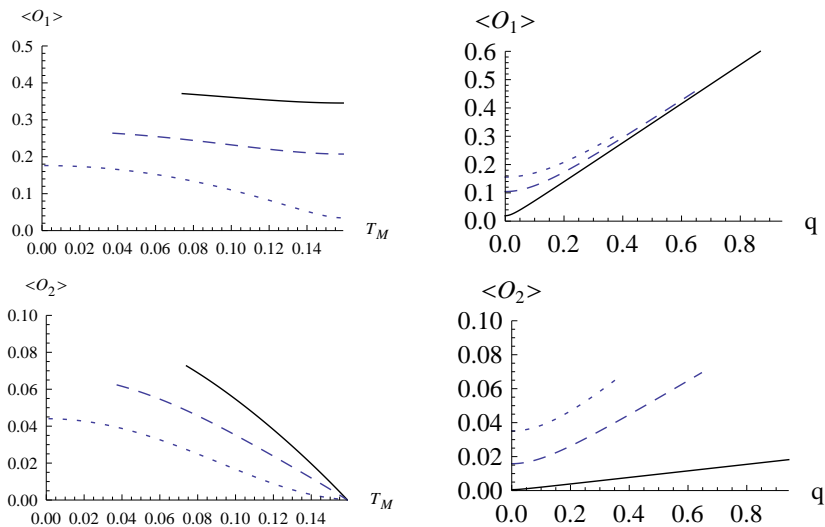}
\caption{The condensates $<{\cal O}_1>$ and $<{\cal O}_2>$ as
functions of temperature $T_M$ and charge density $q$. Left column:
condensates as functions of $T_M$ are plotted for $q=0.5$ (solid),
$0.3$ (dashed) and $0.05$ (dotted). Right column: condensates as
functions of $q$ are shown  for $T_M=0.15$ (solid), $0.1$ (dashed)
and $0.05$ (dotted).} \label{fig3}
\end{figure}
Moreover, near the critical temperature, two condensate behave like
\begin{eqnarray}    \label{conatcri}
\left< {\cal O}_1 \right> &\simeq&  \sqrt{\frac{3}{2 \pi}} \ q, \nonumber \\
\left< {\cal O}_2 \right> &\simeq& \frac{\sqrt{3} \ q}{4 \sqrt{2} \
\pi^{3/2} T_0} \left( 1 - \frac{T_M}{T_0} \right) .
\end{eqnarray}
Comparing these critical exponents with those of the MTZ black
hole~\cite{KPS2}, the black hole charge $q$ which corresponds to the
charge  density on the boundary  field theory decreases the critical
exponents $c_i$ defined by $\left< {\cal O}_i \right> \simeq \left(
1 - \frac{T_M}{T_0} \right)^{c_{i}}$: $c_1=1 \to 0$ and $c_2=2 \to
1$. In Fig. 3, we plot two condensates (\ref{twocond1}) and
(\ref{twocond2}) as functions of $T_M$ and  $q$. As shown in these
figures, we find that the condensate decreases as the temperature
increases. On the other hand, condensates increase as the charge
density  $q$ increases.  At the critical temperature, $\left< {\cal
O}_2 \right>$ goes to zero irrespective of  the charge density,
whereas  $\left< {\cal O}_1 \right>$ remains as a constant value for
nonzero $q$. From this observation and (\ref{conatcri}), we find
that the charge density $q$ significantly changes the condensates.

Now we are in a position to investigate the electric conductivity.
For this purpose, we perform the electromagnetic perturbations
around the CMTZ hyperbolically symmetric  background only by
considering
\begin{equation}
A_\mu=\bar{A}_\mu+a_\mu.
\end{equation}
 Linearizing (\ref{eq13}), a s-wave gauge field fluctuation
satisfies   \beq \label{eqfluct} f_M
\partial_r \left( f_M \partial_r a_w \right) + \left( w^2 - p^2
\left| \bar{\phi} \right|^2 f_M \right) a_w =0, \eeq where  $p$ is a
charge of the complex scalar field and we use the fourier
transformation to define  $a_w$  \beq a_{\theta} (t,r) = \int
\frac{d w}{2 \pi} e^{- i w t} a_w (r). \eeq In deriving
(\ref{eqfluct}), we used the wave function of the lowest harmonic
which corresponds to the lowest eigenvalue in the compact hyperbolic
space $\Sigma_{k=-1}=H^2/\Gamma$ given by $\xi^2+1/4=0$.  We note
that the gauge perturbation in $\varphi$-direction $a_{\varphi}
(t,r)$ can be also described by (\ref{eqfluct}) because we are
working in the hyperbolically symmetric background.  We could not
solve (\ref{eqfluct}) analytically. We may solve this equation for
weak coupling $p^2$ using the perturbation theory. In order to get a
solution near horizon, we introduce the tortoise  coordinate as \beq
r^* = - \int_{r}^{\infty} \frac{d \tilde{r}}{f_M (\tilde{r})} , \eeq
which satisfies \beq \label{relrr*} \frac{d r^*}{d r} =
\frac{1}{f_M}. \eeq  Then, Eq. (\ref{eqfluct}) leads  to the
Schr\"odinger-type equation \beq \frac{d^2a_w}{dr^{*2}}  + \Big( w^2
- p^2 \left| \bar{\phi} \right|^2 f_M \Big) a_w=0. \eeq   Since
$f_M$ is nearly zero near the horizon, we can easily solve the above
equation.  An incoming
 wave near the horizon  is given by
\beq \label{solr*} a^{\leftarrow}_w = e^{- i w r^*} . \eeq   For
small $p^2$ regime, one  keeps the first order perturbation
expansion of $p^2$.   We find that near the boundary, $a_w$ is given
approximately by \beq a_w = e^{- i w r^*} + \frac{p^2}{2 i w} e^{i w
r^*} \int_{r_+}^r d \tilde{r} \left| \bar{\phi} \right|^2 e^{- 2 i w
r^*} - \frac{p^2}{2 i w} e^{- i w r^*} \int_{r_+}^r d \tilde{r}
\left|\bar{ \phi} \right|^2 + \cdots. \eeq  Expanding $a_w$ in the
large $r$, we obtain \beq a_w = a^{(0)}_w + \frac{a^{(1)}_w}{r} +
\cdots \eeq with
\begin{eqnarray}
a^{(0)}_w &=& 1 + \frac{p^2}{2 i w} \int_{r_+}^{\infty} d \tilde{r}
\left| \bar{\phi} \right|^2 \left( e^{- 2 i w r^*} - 1\right), \nonumber\\
a^{(1)}_w &=& i w - \frac{p^2}{2} \int_{r_+}^{\infty} d \tilde{r}
\left|\bar{ \phi} \right|^2 \left( e^{- 2 i w r^*} + 1\right).
\end{eqnarray}
 Then, we read off the conductivity to the first order in $p^2$
as  \beq \sigma (w) = \frac{a^{(1)}_w}{i w a^{(0)}_w} =  1 -
\frac{p^2}{i w} \int_{r_+}^{\infty} d r \left| \bar{\phi} \right|^2
e^{- 2 i w r^*} . \eeq  We note  that the superfluid density $n_s$
is given by the coefficient of the delta function of the real part
of conductivity~\cite{KPS2} \beq Re[\sigma(w)] \sim \pi n_s
\delta(w) , \eeq which is also the coefficient of the pole in the
imaginary part \beq Im[\sigma(w)] \sim  \frac{n_s}{w} \quad {\rm
for} \ w \to 0. \eeq  Then, $n_s$ becomes \beq n_s = \frac{p^2 }{2
\alpha l^2} \frac{G^2 \mu^2}{r_+ + G \mu} . \eeq  Using the
temperature $T_M$ and density $q$, the superfluid density for small
$p^2$  can be rewritten as \beq n^\alpha_s = \frac{6 p^2 T_0^3}{
(T_0 + T_M)^2} \left( q^2 + \frac{\left( T_0^2 -  T_M^2
\right)^2}{64 \pi^2 G T_0^6} \right). \eeq In the limit of $q^2 \to
0$, we find that $n^q_s$ reduces to $n_s$ of the probe limit in
Ref.~\cite{KPS2}. Also, we note that near the zero temperature
$T_M=0$,
\begin{equation}
n^q_s(0)-n^q_s(T_M) \simeq T_M \end{equation} which matches the low
temperature behavior of heat capacity in (\ref{TCM}).

\section{Discussion}

First of all, we would like to mention that the exact gravity dual
of a gapless superconductor based on the TBH-MTZ black
holes~\cite{KPS2} was considered as a probe limit-description of
holographic superconductor  because these black hole belongs to
neutral black holes without charge.   We have studied the CMTZ black
hole as a model of analytic hairy black hole for holographic
superconductor.  This situation is opposite to a case of numerical
hairy black holes that the charged black holes without scalar hair
were known explicitly, while the charged black holes with scalar
hair were found numerically.  In this work,  we have reanalyzed the
phase transition between HRNAdS and CMTZ black holes. However, this
transition unlikely occur, which means that HRNAdS black holes may
not be  the corresponding black hole without scalar hair. We note
that a special transition between  HRNAdS with
$\tilde{q}=i\sqrt{G}\mu$ and CMTZ black hole with $q=i\sqrt{G}\mu$
\cite{KPS1} is not allowed because this channel is beyond the real
transition.  Explicitly, the presence of black hole charge affects
thermodynamics of CMTZ black holes drastically: changing from
$\alpha=\frac{2\pi G}{3l^2}$ to $\alpha=0$ means increasing of
charge from $q=0$ to $q=\infty$.  This is mainly because there
exists some restriction on defining the  entropy and  free energy of
CMTZ black hole due to the black hole charge $q$. The onset of black
hole charge disturbs a transition to the corresponding black hole
without scalar hair.  An exceptional case is a pair of neutral black
holes of MTZ and TBH where the second order phase transition was
well established~\cite{Siop,Myungsh}.

As an analytic treatment for holographic superconductor, we have
developed superconductor in the bulk and superfluidity on the
boundary using the CMTZ black hole below the critical temperature.
This might provide an analytic hairy black hole to study the
holographic superconductor.  It was shown that the presence of
charge density destroys the condensates, which is in accord  with
the thermodynamic analysis of the CMTZ black hole.
 At the
critical temperature $T=T_0$, $\left< {\cal O}_2 \right>$ goes to
zero, irrespective of  the charge  density $q$,  whereas  $\left<
{\cal O}_1 \right>$ remains as a constant value for nonzero $q$.
Hence, we find  that the charge density $q$ affects the onset of the
superconductivity.

Finally, we consider the zero temperature limit of holographic
superconductor~\cite{LNDP}. As was shown in black hole
thermodynamics, each allowed region for free energy excludes the
extremal black hole at the zero temperature $T_M(r_e)=0$ except the
MTZ  black hole. The entropies of extremal CMTZ black holes all are
negative (see (\ref{extremalE})) which is unacceptable for black
hole thermodynamics, while the entropy of extremal MTZ black hole is
zero~\cite{HRZ}.
 We observe that similar
results for the conductivity depending on the charge  density were
found in the Horowitz-Roberts model \cite{HRZ} and Ref.\cite{KZ}.

In conclusion, we have shown  that the onset of charge density on
the boundary disturbs making a solid superfluidity phase by studying
an analytic hairy solution  of CMTZ black hole. Hence, it seems
difficult to find an exact hairy black hole which describes the
holographic superconductor and its corresponding black hole without
scalar hair, compared to numerical hairy black holes.

\section*{Acknowledgments}
Y. Myung was in part  supported  by Basic Science Research Program
through the National Research  Foundation (NRF) of Korea funded by
the Ministry of Education, Science and Technology (2011-0027293) and
the NRF grant funded by the Korea government (MEST) through the
Center for Quantum Spacetime (CQUeST) of Sogang University with
grant number 2005-0049409. C. Park was supported by the NRF grant
funded by the Korea government (MEST) through the Center for Quantum
Spacetime (CQUeST) of Sogang University with grant number
2005-0049409.


\begin{thebibliography}{99}

\bibitem{Gubs}
  S.~S.~Gubser,
  Phys.\ Rev.\  D {\bf 78}, 065034 (2008)
  [arXiv:0801.2977 [hep-th]].


\bibitem{HHH}
  S.~A.~Hartnoll, C.~P.~Herzog and G.~T.~Horowitz,
  Phys.\ Rev.\ Lett.\  {\bf 101}, 031601 (2008)
  [arXiv:0803.3295 [hep-th]].

\bibitem{HartC}
  S.~A.~Hartnoll,
  Class.\ Quant.\ Grav.\  {\bf 26}, 224002 (2009)
  [arXiv:0903.3246 [hep-th]].

\bibitem{Horo}
  G.~T.~Horowitz,
  arXiv:1002.1722 [hep-th].



\bibitem{HerzJ}
  C.~P.~Herzog,
  J.\ Phys.\ A  {\bf 42}, 343001 (2009)
  [arXiv:0904.1975 [hep-th]].


  \bibitem{MFK}
  K.~Maeda, S.~Fujii and J.~i.~Koga,
  Phys.\ Rev.\  D {\bf 81}, 124020 (2010)
  [arXiv:1003.2689 [gr-qc]].


\bibitem{BHMS}
  P.~Basu, J.~He, A.~Mukherjee and H.~H.~Shieh,
  JHEP {\bf 0911}, 070 (2009)
  [arXiv:0810.3970 [hep-th]].

\bibitem{HPu}
  C.~P.~Herzog and S.~S.~Pufu,
  JHEP {\bf 0904}, 126 (2009)
  [arXiv:0902.0409 [hep-th]].







\bibitem{KPS2}
  G.~Koutsoumbas, E.~Papantonopoulos and G.~Siopsis,
  JHEP {\bf 0907}, 026 (2009)
  [arXiv:0902.0733 [hep-th]].

\bibitem{MTZ}
  C.~Martinez, R.~Troncoso and J.~Zanelli,
  Phys.\ Rev.\  D {\bf 70}, 084035 (2004).


 \bibitem{Siop}
  G.~Koutsoumbas, S.~Musiri, E.~Papantonopoulos and G.~Siopsis,
  JHEP {\bf 0610}, 006 (2006)
  [arXiv:hep-th/0606096].




  \bibitem{Myungsh}
  Y.~S.~Myung,
  Phys.\ Lett.\  B {\bf 663}, 111 (2008)
  [arXiv:0801.2434 [hep-th]].


\bibitem{HerzA}
  C.~P.~Herzog,
  Phys.\ Rev.\  D {\bf 81}, 126009 (2010)
  [arXiv:1003.3278 [hep-th]].



  \bibitem{BFB}
  P.~Breitenlohner and D.~Z.~Freedman,
  Annals Phys.\  {\bf 144}, 249 (1982).


\bibitem{CMTZ}
  C.~Martinez, J.~P.~Staforelli and R.~Troncoso,
  Phys.\ Rev.\  D {\bf 74}, 044028 (2006)
  [arXiv:hep-th/0512022].

\bibitem{KPS1}
  G.~Koutsoumbas, E.~Papantonopoulos and G.~Siopsis,
  JHEP {\bf 0805}, 107 (2008)
  [arXiv:0801.4921 [hep-th]].

 \bibitem{Lemos}
  J.~P.~S.~Lemos,
  Phys.\ Lett.\  B {\bf 353}, 46 (1995).
\bibitem{Vanzo}
  L.~Vanzo,
  Phys.\ Rev.\  D {\bf 56}, 6475 (1997).

\bibitem{BLP}
  D.~R.~Brill, J.~Louko and P.~Peldan,
  Phys.\ Rev.\  D {\bf 56}, 3600 (1997).

 \bibitem{myungsds}
  Y.~S.~Myung,
  Phys.\ Lett.\  B {\bf 645}, 369 (2007).

\bibitem{BM}
  D.~Birmingham and S.~Mokhtari,
  Phys.\ Rev.\  D {\bf 76}, 124039 (2007)
  [arXiv:0709.2388 [hep-th]].

\bibitem{NVZ}
  M.~Nadalini, L.~Vanzo and S.~Zerbini,
  Phys.\ Rev.\  D {\bf 77}, 024047  (2008)
  [arXiv:0710.2474 [hep-th]].

\bibitem{HP} S. W. Hawking  and  D. N. Page, Commun. Math. Phys. {\bf 87},
577 (1983).

\bibitem{MKP}
  Y.~S.~Myung, Y.~W.~Kim and Y.~J.~Park,
  Phys.\ Rev.\  D {\bf 78}, 084002 (2008)
  [arXiv:0805.0187 [gr-qc]].



\bibitem{LNDP}
  B.~H.~Lee, S.~Nam, D.~W.~Pang and C.~Park,
  Phys.\ Rev.\  D {\bf 83} (2011) 066005
  [arXiv:1006.0779 [hep-th]].



\bibitem{HRZ}
  G.~T.~Horowitz and M.~M.~Roberts,
  JHEP {\bf 0911}, 015 (2009)
  [arXiv:0908.3677 [hep-th]].

\bibitem{KZ}
  R.~A.~Konoplya and A.~Zhidenko,
  Phys.\ Lett.\  B {\bf 686}, 199 (2010)
  [arXiv:0909.2138 [hep-th]].



\end{thebibliography}
\end{document}